\documentclass[times,twocolumn,final]{elsarticle}

\usepackage{framed,multirow}

\usepackage{amssymb}
\usepackage{amsmath}
\usepackage{latexsym}
\usepackage{mathtools}
\usepackage{subcaption}

\usepackage{url}
\usepackage{xcolor}

\usepackage{hyperref}
\usepackage{siunitx}
\DeclareSIUnit\Molar{\textsc{m}}

\definecolor{newcolor}{rgb}{.8,.349,.1}
\definecolor{ForestGreen}{RGB}{34,139,34}

\journal{Medical Image Analysis}

\begin{document}

\begin{frontmatter}

\title{Physics-informed neural networks for myocardial perfusion MRI quantification}

\author[1]{Rudolf L.M. {van Herten}}
\author[2]{Amedeo {Chiribiri}}
\author[1,3]{Marcel {Breeuwer}}
\author[1]{Mitko {Veta}}
\author[2]{Cian M. {Scannell}\corref{cor1}}
\cortext[cor1]{Corresponding author. \\ 
		\textit{Email address}: \href{mailto:cian.scannell@kcl.ac.uk.}{cian.scannell@kcl.ac.uk.} (C.M. Scannell)}

\address[1]{Department of Biomedical Engineering, Medical Image Analysis group, Eindhoven University of Technology, Eindhoven, The Netherlands}
\address[2]{School of Biomedical Engineering and Imaging Sciences, King’s College London, United Kingdom}
\address[3]{Philips Healthcare, Best, The Netherlands}

\begin{abstract}
Tracer-kinetic models allow for the quantification of kinetic parameters such as blood flow from dynamic contrast-enhanced magnetic resonance (MR) images. Fitting the observed data with multi-compartment exchange models is desirable, as they are physiologically plausible and resolve directly for blood flow and microvascular function. However, the reliability of model fitting is limited by the low signal-to-noise ratio, temporal resolution, and acquisition length. This may result in inaccurate parameter estimates.

This study introduces physics-informed neural networks (PINNs) as a means to perform myocardial perfusion MR quantification, which provides a versatile scheme for the inference of kinetic parameters. These neural networks can be trained to fit the observed perfusion MR data while respecting the underlying physical conservation laws described by a multi-compartment exchange model. Here, we provide a framework for the implementation of PINNs in myocardial perfusion MR.

The approach is validated both \textit{in silico} and \textit{in vivo}. In the \textit{in silico} study, an overall reduction in mean-squared error with the ground-truth parameters was observed compared to a standard non-linear least squares fitting approach. The \textit{in vivo} study demonstrates that the method produces parameter values comparable to those previously found in literature, as well as providing parameter maps which match the clinical diagnosis of patients.
\end{abstract}

\begin{keyword}
Physics informed neural networks\sep Small data\sep Tracer-kinetic modelling\sep Myocardial perfusion MRI
\end{keyword}

\end{frontmatter}


\section{Introduction}

The evaluation of dynamic contrast-enhanced magnetic resonance imaging (DCE-MRI) has proven to be a popular tool for the assessment of tissue physiology in various diseases \cite{Nagel2003, Turnbull2009, Heye2014}. In particular, stress perfusion cardiac magnetic resonance (CMR) is becoming an established technique for the non-invasive assessment of patients with suspected coronary artery disease (CAD) \cite{Montalescot2013}. A series of large trials have demonstrated its efficacy in clinical practice \cite{Greenwood2012, Schwitter2013, Greenwood2016}, it has recently been shown to be non-inferior to the invasive reference standard, fractional flow reserve (FFR), for the management of patients with suspected CAD \cite{Nagel2019} and to be cost-effective \cite{Kwong2019}. However, since the images are difficult to interpret and ischaemic burden may be underestimated for patients suffering from multi-vessel disease, visual assessment of the severity of CAD using stress perfusion CMR remains limited \cite{Villa2018}. As such a quantitative analysis of perfusion is proposed as a reproducible and user-independent alternative to the visual assessment \cite{Patel2010}.

The quantification of myocardial blood flow (MBF) has shown promising diagnostic accuracy and prognostic value \cite{Lockie2010, Hsu2018, Sammut2017} and has the potential to allow more widespread clinical adoption of stress perfusion CMR. The basis for myocardial perfusion quantification is the use of tracer-kinetic modelling, which provides a relation between the concentration-time curves derived from DCE-MRI, and patient-specific physiological parameters, such as MBF \cite{Sourbron2011}. In the case of myocardial perfusion this is achieved by modelling how the contrast agent passes from the left ventricle (LV) into the myocardium, allowing for the inference of kinetic parameters.

The identification of the kinetic parameters is not, however, a trivial task. Several studies have shown that reliable quantification of myocardial perfusion was only achieved with relatively simple models such as Fermi-constrained deconvolution, but not with multi-compartment exchange models \cite{Broadbent2013, Schwab2015, Likhite2017}. One of the underlying reasons for the difficulties in model fitting is that non-linear regression problems tend to get stuck in local optima \cite{Kelm2009, Dikaios2017}. Though the model-based concentration curves may match the noisy observed data, the inferred parameters can be far from the actual values. It has further been shown that model parameters are correlated \cite{Romain2017}, which causes several distinct parameter combinations to produce concentration curves which may all very well fit the observed data, therefore making the identification of the one true set of parameters difficult \cite{Buckley2002, Ahearn2005}. This has lead to more complex fitting algorithms being proposed but these have not yet seen widespread adoption \cite{Kelm2009, Dikaios2017, Scannell2020, Dikaios2020}.

Given the limitations of the currently used methodologies, this study introduces a novel class of algorithms for the quantification of myocardial perfusion, \textit{physics-informed neural networks} (PINNs) \cite{Raissi2019}. PINNs are based on the universal approximation theorem of neural networks which is leveraged to solve supervised learning tasks while respecting the given physical laws in terms of ordinary or partial differential equations \cite{Hornik1989}. Specifically, the solutions to the tracer-kinetic model differential equations are approximated by neural networks which are trained to produce outputs that both fit the available data, and satisfy the underlying physical conservation laws.

\section{Methods}
\begin{figure*}[t]
	\centering
	\includegraphics[width = 0.96\linewidth]{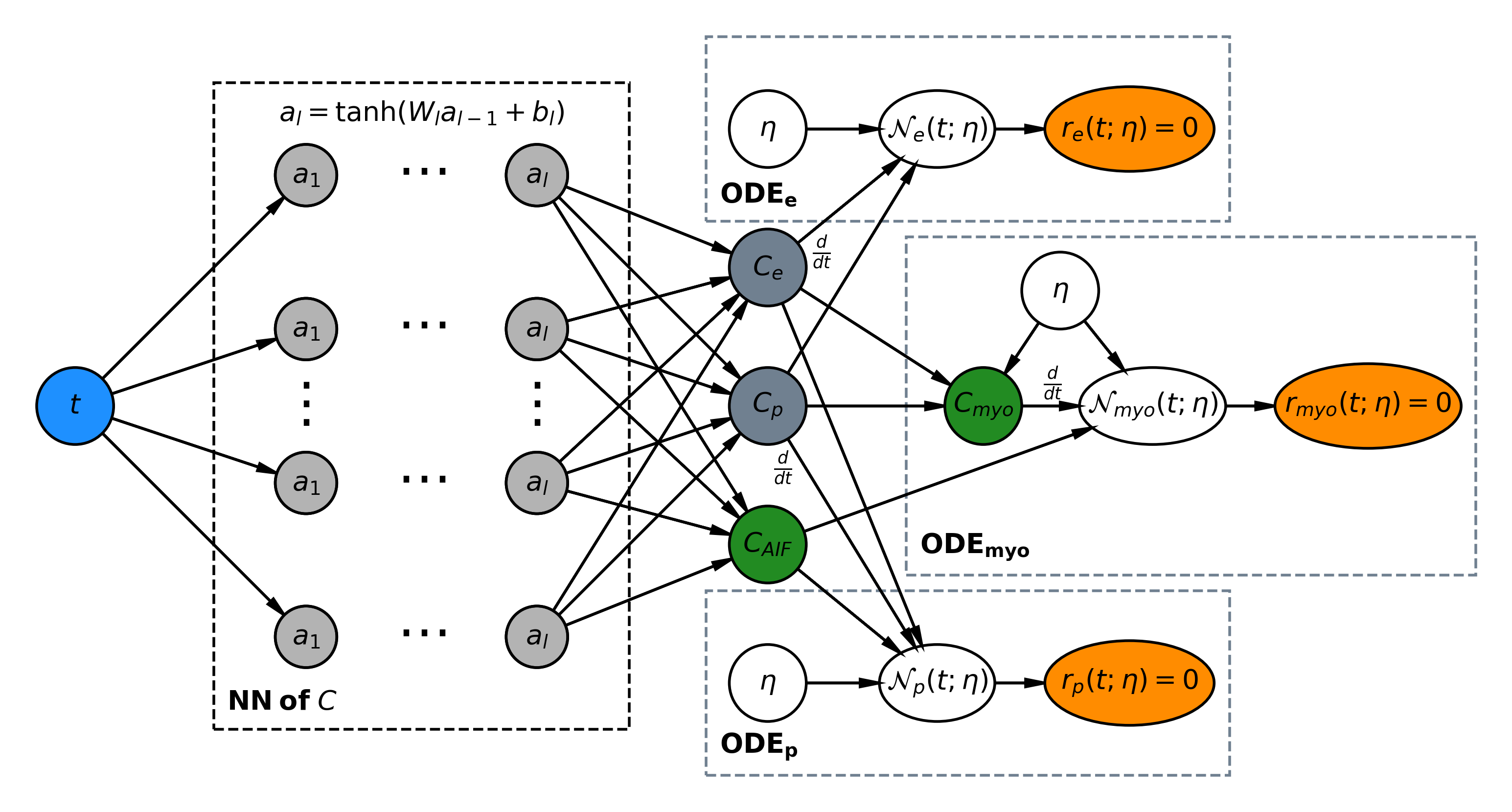}
	\caption{A schematic representation of the physics-informed neural network. The green coloured nodes represent values which can be fit directly with contrast agent measurements, thereby regulating the neural network. Derivatives of these measurement approximations are then combined with the linear operator of each ODE, described by $\mathcal{N}$. Orange coloured nodes directly denote the residuals of the differential equations. Loss functions $\mathcal{L}_{C}$ and $\mathcal{L}_{r}$ can then be defined to optimise the parameters of the neural networks. $\mathcal{L}_{C}$ minimises the difference between the 
	network output concentrations and the observed MRI concentrations while $\mathcal{L}_{r}$, minimises the residuals of the ODEs. The set of 2CXM parameters $\{F_p, v_p, v_e, PS\}$ are denoted by the $\eta$ nodes in the figure.}
	\label{MyoPINN}
\end{figure*}
\subsection{Tracer-kinetic modelling}
Tracer-kinetic models present a mathematical description for the physics which underlie the transport process of a contrast agent across a tissue \cite{Ingrisch2013}. The two-compartment exchange model (2CXM) has been suggested as appropriate for modelling myocardial perfusion \cite{Jerosch-Herold2010}. In this case, the perfusion unit (represented by a single pixel in DCE-MRI) is modelled as a system of two interacting compartments, the plasma and the interstitium. This gives rise to a pair of coupled ordinary differential equations (ODEs) which describe the evolution of the concentration of contrast agent over time:
\begin{equation}
v_p\frac{dC_p(t)}{dt} = F_p(C_{AIF}(t) - C_p(t)) + PS(C_e(t) - C_p(t)),
\label{ODEp}
\end{equation}
\begin{equation}
v_e\frac{dC_e(t)}{dt} = PS(C_p(t) - C_e(t)).
\label{ODEe}
\end{equation}
Here, $C_p(t)$ and $C_e(t)$ are the concentration of contrast agent, gadolinium [Gd] in the plasma and interstitial space at time t, respectively (in units of molarity (\si{\Molar})). $C_{AIF}(t)$, the arterial input function (AIF), is the assumed input to the system that is being modelled (also in \si{\Molar}). In myocardial perfusion quantification this is sampled from the LV. $F_p$ is the plasma flow (\si{\milli\litre\per\min\per\milli\litre}), $v_p$ is the fractional plasma volume (dimensionless), $v_e$  is the fractional interstitial volume (dimensionless) and PS is the permeability-surface area product (\si{\milli\litre\per\min\per\milli\litre}). A weighted sum of the concentration within these separate compartments then produces a representation for the concentration within the myocardial tissue ($C_{myo}(t)$ in \si{\Molar}):

\begin{equation}
C_{myo}(t) = v_pC_p(t) + v_eC_e(t).
\label{Cmyo_eq}
\end{equation}
This is fit to the observed imaging data to infer the parameters:  $F_p, v_p, v_e$, and $PS$. 

\subsection{Physics-informed neural networks}
PINNs are a new framework within the deep learning paradigm which employ deep neural networks to approximate the solution to physical systems, primarily ordinary and partial differential equations \cite{Raissi2019}. With this framework, the parameters of the neural network are constrained and learned in two ways. Firstly, the direct outputs of the neural network are trained to fit the observed data, and secondly, the neural network is constrained to satisfy the underlying physical laws that govern this observed data. In this particular case, the physical laws are modelled by the ODEs derived from the 2CXM, found in Eq. \ref{ODEp} and \ref{ODEe}.

As such, a neural network $f(t; \theta)$ is defined that approximates the solution to these equations with parameters $\theta$ for all pixels $k$ in an imaging slice. This is the mapping from time $t$ to the solutions of the ODEs at time $t$:
\begin{equation}
t \xmapsto{f_\theta} \begin{bmatrix}
C_{p}^1(t)  \\[2pt]
C_{p}^2(t)  \\[2pt]
\vdots      \\[2pt]
C_{p}^K(t)
\end{bmatrix}, \begin{bmatrix}
C_{e}^1(t)  \\[2pt]
C_{e}^2(t)  \\[2pt]
\vdots      \\[2pt]
C_{e}^K(t)
\end{bmatrix}, C_{AIF}(t).
\label{NN_mapping}
\end{equation}
The outputs of the neural network are also constrained by an additional loss function derived from the set of residuals corresponding to the 2CXM:
\begin{equation}
\begin{split}
r_{p}(t) & := v_p\frac{dC_p}{dt}  - PS(C_e - C_p) - F_p(C_{AIF} - C_p),\\
r_{e}(t) & := v_e\frac{dC_e}{dt} - PS(C_p - C_e).\\
\end{split}
\label{Residuals}
\end{equation}
As well as training the parameters of the neural network, these residuals may then also be used to learn the kinetic parameters present in the ODEs. This ODE residual loss encourages the PINN to produce physically plausible results and is made possible by the automatic differentiation functionality of deep learning frameworks. A schematic representation of the PINN is shown in Fig. \ref{MyoPINN}.

The loss function for training the neural network is then given as the sum over all pixels $j$ of the weighted sum of these ODE residuals ($\mathcal{L}_{r}$), the sum of squared differences between the predicted and the observed concentrations ($\mathcal{L}_{C}$), the initial conditions ($\mathcal{L}_{b}$) which enforce that there is not contrast in the system at time $t=0$), and a regularisation term to enforce non-negative concentration values ($\mathcal{L}_{reg}$):

\begin{equation}
	\begin{split}
		\mathcal{L} &= \frac{1}{K}\sum\limits_{j=1}^{K}( w_r\mathcal{L}_{r}^j + w_C\mathcal{L}_{C}^j + w_b\mathcal{L}_{b}^j + w_{reg}\mathcal{L}_{reg}^j).
	\end{split}
	\label{Combined_loss}
\end{equation}
$w_C$, $w_r$, $w_b$, and $w_{reg}$ are the weightings for each term in the combined loss, $K$ is the total number of pixels, and the exact form of each term in Equation \ref{Combined_loss} is given in Appendix A.

\subsection{Reduced form 2CXM} 
In this work, the idea of constraining the residuals of the total myocardial concentration rather than the plasma and interstitial compartment concentrations (as in Eq. \ref{ODEp} and \ref{ODEe}) will be tested. This can be achieved by taking the time derivative of Eq. (\ref{Cmyo_eq}) and replacing its terms with the ODEs of the 2CXM found in Eq. (\ref{ODEp}) and (\ref{ODEe}):
\begin{equation}
\begin{split}
\frac{dC_{myo}(t)}{dt} & = v_p\frac{dC_p(t)}{dt} + v_e\frac{dC_e(t)}{dt} \\
& = F_p(C_{AIF}(t) - C_p(t)).
\end{split}
\label{ODEmyo}
\end{equation}
This ODE is further rewritten into a residual function, that can be incorporated into the PINN:
\begin{equation}
r_{myo}(t) := \frac{dC_{myo}}{dt} - F_p(C_{AIF} - C_p).\\
\label{Residuals_reduced}
\end{equation}

\section{Experiments} 
The proposed algorithm will be tested in two separate ways on both simulated and patient data, allowing for a qualitative and quantitative assessment regarding the benefits of the methodology compared to the currently used non-linear least squares (NLLS) solution.

\subsection{Generation of the digital phantom}
\label{Dro_study}
In the first case a 2CXM digital reference object (DRO) is constructed, for which the pixelwise parameters $\eta$ are known. 144 myocardial tissue concentration curves with different parameter combinations of $F_p$, $v_p$, $v_e$, and $PS$ are subsequently simulated using a gamma-variate function for the AIF, producing the data which the model should fit. This is based on the DRO proposed by Debus et al. \cite{Debus2019} with the tissue kinetics, AIF, and time resolution adapted to more realistically represent myocardial perfusion. The 2CXM parameters used for this study are therefore as follows:
\begin{equation}
\begin{split}
F_p & \in \{0.5, 1.0, 1.5, 2.0\}, \quad v_p \in \{0.02, 0.05, 0.1, 0.2\} \\
v_e & \in \{0.1, 0.2, 0.5\}, \qquad \hspace{2pt} PS \in \{0.5, 1.5, 2.5\}.
\end{split}
\label{DRO_params}
\end{equation}
This DRO gives rise to a volume of time curves with spatial dimensions $40 \times 120 \times 3$ and a unique subset of 2CXM parameters for every $10 \times 10 \times 1$ block of pixels. This is simulated at a temporal resolution of 0.02 min over a time span of $T = $ 2 \si{min}, creating tissue curves with a total of $N_C$ = 100 time points as a result. Random normal noise was subsequently added to the generated curves such that the final curve has a signal-to-noise ratio of 17.5.

\subsection{Digital phantom study}
Using the DRO, a comparative study of five separate methods is performed, each with the goal of estimating the 2CXM parameters $\eta$ given the generated noisy data. 
\begin{enumerate}
	\item The 2CXM PINN in which the original two residual functions provided in Eq. \ref{Residuals} are used in conjunction with the described PINN.
	\item The 2CXM + Mesh PINN which builds on the 2CXM PINN and in addition to temporal information, the spatial coordinates of the data are provided as input to the model, to regularise the neural network to produce similar concentration approximations for similar coordinates. 
	\item The Reduced 2CXM PINN which solely relies on the reduced-form ODE found in Eq. \ref{Residuals_reduced}.
	\item The Combined PINN which jointly optimises the 2CXM and the Reduced 2CXM residuals.
	\item The standard NLLS optimisation method.
\end{enumerate}
For each of these methods, the normalised mean square error (NMSE) between the true and estimated kinetic parameters is reported along with a structural similarity index (SSIM) to test the overall structural coherence as compared to the ground truth (GT).

\subsection{Patient data study}
The proposed method was tested on clinical stress perfusion scans for 8 patients. The study was conducted in accordance with the Declaration of Helsinki (2000) and was approved by the National Research Ethics Service (15/NS/0030). All patients provided written informed consent. The patient datasets used for this project comprise of examinations performed on a 3T system (Achieva TX, Philips Healthcare, Best, The Netherlands), with a 32-channel cardiac phased array receiver coil. For each patient, a total of three LV short-axis slices were attained at the apical, mid-cavity, and basal level. These slices were acquired at mid-expiration with a saturation-recovery gradient echo method. Stress images were obtained during adenosine-induced hyperaemia, and for each acquisition 0.075 mmol/kg of bodyweight gadolinium contrast agent was administered at 4 mL/s, followed by 20 mL saline flush per acquisition. Each bolus of gadobutrol was preceded by a diluted pre-bolus with 10\% of the dose in order to mitigate the non-linear relationship between MR intensity values and contrast agent concentration \cite{Ishida2011}. 

All acquired perfusion images were subsequently corrected for respiratory motion and processed end-to-end through a deep learning-based pipeline \cite{Scannell2019, Scannell2020a}. Pixelwise signal-intensity curves were extracted from the myocardial segmentation, and were split into time intervals corresponding to the pre-bolus injection and the main bolus injection in order to perform quantification. Finally, the American heart association (AHA) representation was used to assign the pixelwise parameter estimates to AHA segments through automatically computed right ventricular insertion points \cite{Cerqueira2002}. The haematocrit value (HCT) was assumed to be 0.45 and the specific density of the myocardium was assumed to be 1.05 \si{\gram\per\milli\litre}. These were used to convert plasma flow and volume ($F_p$ and $v_p$) to blood flow and volume ($F_b$ and $v_b$). 

The kinetic parameters of these patients are estimated using the best model found in the above Digital phantom study. Qualitative assessment of the stress perfusion scans was also conducted by an experienced Level-3 cardiologist (A.C) \cite{Plein2011}. The distribution of the MBF values in AHA segments which were visually positive for ischaemia was compared to the MBF values in segments negative for ischaemia using the nonparametric Mann-Whitney U test. The significance level $\alpha=0.05$ was used to determine statistical significance. This analysis was performed using SciPy \cite{Virtanen2020}.

\subsection{Optimisation details}
The optimisation of the physiological parameters of the 2CXM is performed in log-space to encourage the estimation of positive values. These parameters are further initialised to a value range which is within the physiological range. The loss function is also adapted to penalise predictions of negative concentration values. While the loss on the observed measurements is enforced at the available observed measurements, the residual solution is enforced at a total of $N_r$ = 500 points generated through a random uniform distribution within the time domain. The ability to enforce the residuals at these collocation points is the reason why the AIF is also predicted by the neural network, as in Equation \ref{NN_mapping}.

The neural network weights are initialised using Glorot uniform initialisation and are iteratively updated using the Adam optimisation algorithm with a learning rate of 0.001 \cite{Glorot2010, Kingma2015}. The network uses fully-connected layers and consists of two hidden layers with 32 units each, which are all followed by a hyperbolic tangent (tanh) activation function and a batch normalisation layer. This choice of activation function is justified by the choice of relatively shallow neural networks commonly employed, and the importance of balanced gradient flow required \cite{LeCun2012}. Models are trained for 25000 iterations. The architecture was inspired by previous work in the field \cite{Raissi2019} and there was no attempt to optimise this. The number of iterations was chosen empirically along with the weight factors $w_C = 5$, $w_r = 1$, $w_b = 1$, and $w_{reg} = 1$. All aforementioned methods and optimisations are implemented with the Tensorflow 2 deep learning library \cite{Abadi2016}.

As is common for deep learning, the input and output data are normalised in order to stabilise the training process \cite{Glorot2010}. The input data is standardised as follows:
\begin{equation}
\hat{t} = \frac{t - \mu_t}{\sigma_t},
\label{t_normalisation}
\end{equation}
where $\mu_t$ and $\sigma_t$ are the mean and standard deviation of the temporal coordinates respectively. Secondly, the output concentration values are normalised to the range $[0, 1]$ in order to further mitigate any scaling performed by the neural network itself:
\begin{equation}
\hat{C}_{myo} = \frac{C_{myo}}{\max{(C_{AIF}})}, \quad \hat{C}_{AIF} = \frac{C_{AIF}}{\max{(C_{AIF}})}.
\label{c_normalisation}
\end{equation}
The ODEs are re-written to account for these scalings, as shown in the Appendix B.

The open source implementation is provided on github\footnote{\url{https://github.com/cianmscannell/myo_pinn}}, as well as scripts to reproduce the results. All derived data is provided, including the trained model weights, but the raw patient data is not shared for ethical reasons.

\section{Results}
\subsection{Digital phantom study}
A summary of the DRO study performance is presented in table \ref{DRO_results}, showing both the NMSE and SSIM between the estimated parameters and the ground-truth DRO. The overall metrics are shown as well as the values on the level of individual parameters. An example 2D slice inference of the 3D DRO is presented in Fig. \ref{DRO_PS1,5}.  This figure demonstrates a slice taken from the z-axis, along which the PS value varies, therefore producing a figure for which the GT value of the PS parameter is constant ($PS = 1.5$). It is seen that there is a lower overall NMSE for both the 2CXM PINN and the Combined PINN as compared to the standard NLLS approach. Indeed, the Combined PINN is the best performing model overall and for all kinetic parameters except for $F_p$. This model is therefore chosen for use in the patient data study results presented in Section 4.2. 
\begin{table*}
	\centering
	\caption{The NMSE for the individual and combined kinetic parameter estimates, evaluated for each proposed methodology. The SSIM value for the estimates is presented in brackets next to each NMSE score. Green-colored values represent those scores which were the best amongst all models. Scores are based on the digital reference dataset with SNR of 17.5 described in the Methods section.}
	\begin{tabular}{l || c | c | c | c | c | c | c| c | c | c}
		\multicolumn{1}{r}{}& \multicolumn{10}{ c }{Parameter} \\
		\multicolumn{1}{r ||}{} &  \multicolumn{2}{c}{$All$}    & \multicolumn{2}{c}{$F_p$}             & \multicolumn{2}{c}{$v_p$}         & \multicolumn{2}{c}{$v_e$}         & \multicolumn{2}{c}{$PS$}\\
		&  \multicolumn{1}{c }{NMSE}  &  \multicolumn{1}{c |}{SSIM} &  \multicolumn{1}{c }{NMSE}  &  \multicolumn{1}{c |}{SSIM}&  \multicolumn{1}{c }{NMSE}  &  \multicolumn{1}{c |}{SSIM}&  \multicolumn{1}{c }{NMSE}  &  \multicolumn{1}{c |}{SSIM}&  \multicolumn{1}{c }{NMSE}  &  \multicolumn{1}{c }{SSIM}\\
		\hline
		2CXM        & 0.13 & 0.58       & 0.17 & 0.79  & {\color{ForestGreen} \textbf{0.03}} & {\color{ForestGreen} \textbf{0.51}} & {\color{ForestGreen} \textbf{0.01}} & {\color{ForestGreen} \textbf{0.64}} & {\color{ForestGreen} \textbf{0.30}} & 0.39 \\
		2CXM + Mesh    & 0.18 & 0.46 & 0.07 & 0.73 & 0.07 & 0.38 & 0.04 & 0.45 & 0.56 & 0.28\\
		Reduced        & 0.19 & 0.25 & {\color{ForestGreen} \textbf{0.02}} & 0.77 & 0.04 & 0.17 & 0.17 & 0.04 & 0.53 & 0.01 \\
		Combined            & {\color{ForestGreen} \textbf{0.11}} & {\color{ForestGreen} \textbf{0.59}}       & 0.13 & {\color{ForestGreen} \textbf{0.80}}  & {\color{ForestGreen} \textbf{0.03}} & {\color{ForestGreen} \textbf{0.51}} & {\color{ForestGreen} \textbf{0.01}} & {\color{ForestGreen} \textbf{0.64}} & {\color{ForestGreen} \textbf{0.30}} & 0.40 \\
		NLLS            & 0.21 & 0.51 & 0.03 & 0.61 & 0.08 & 0.34 & 0.03 & 0.43 & 0.70 & {\color{ForestGreen} \textbf{0.65}} \\
	\end{tabular}
	\label{DRO_results}
\end{table*}

\subsection{Patient data study}
\begin{figure}[h!]
	\centering
	\includegraphics[width = .47\textwidth]{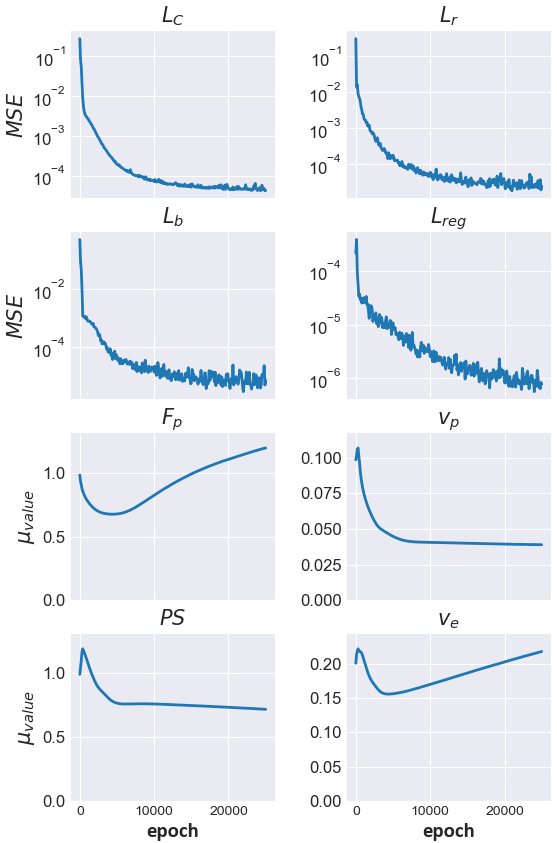}
	\caption{The training curves which show both the evolution of the loss terms (top two rows) and the (mean) estimated parameter values (bottom two rows) over the course of the training.}
	\label{training_curve}
	\vspace{-11pt}
\end{figure}

\begin{figure}[h!]
	\centering
	\includegraphics[width = .47\textwidth]{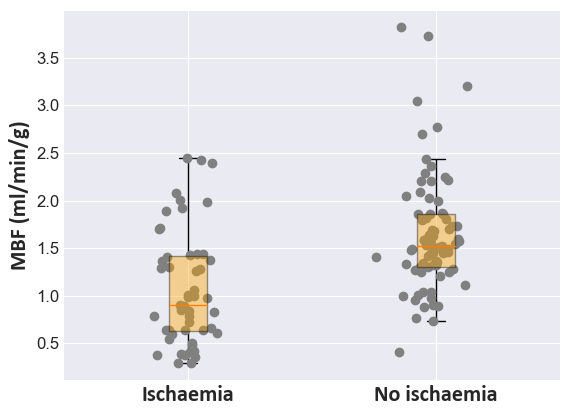}
	\caption{The distribution of inferred MBF values in AHA segments with ischaemia (left) and without ischaemia (right).}
	\label{validation}
	\vspace{-11pt}
\end{figure}

5/8 patients were reported as being visually positive for ischaemia and 3/8 patients were reported as having visually normal images. Table \ref{mean and std} shows the median value (25\textsuperscript{th} percentile, 75\textsuperscript{th} percentile) for all inferred kinetic parameters. These results are broadly in line with values reported in the literature previously \cite{Scannell2020}. Figure \ref{training_curve} shows the evolution of the constituent loss terms and the mean value (over the whole patient) of the estimated kinetic parameters over the training process for a representative image slice (patient 3, basal slice). The estimated MBF maps for two representative patients (one with and one without ischaemia) are shown in Fig. \ref{pt_results} along with the MR images. The corresponding images for all other patients are shown in Appendix C. Homogenous blood flow is shown in the normal case with areas of clearly reduced blood flow seen in the ischaemic patient. The MBF ($F_b$) median value (25\textsuperscript{th} percentile, 75\textsuperscript{th} percentile in AHA segments with ischaemia on the visual assessment was 0.9 (0.62, 1.42) \si{\milli\litre\per\min\per\gram}. This was significantly lower ($p<0.001$) than the equivalent value in normal segments of 1.52 (1.31, 1.86) \si{\milli\litre\per\min\per\gram}. The distribution of these two sets of values is shown in a boxplot in Figure \ref{validation}. 
\begin{table}[h]
	\centering
		\caption{The median value (25\textsuperscript{th} percentile, 75\textsuperscript{th} percentile) of kinetic parameters estimates for the patient data, using the Combined PINN model.}
	\begin{tabular}{l || c  c }
		
		\multicolumn{1}{c}{Parameter} & 
		\multicolumn{1}{c}{Median} & \multicolumn{1}{c}{\vtop{\hbox{\strut (25\textsuperscript{th} percentile,}\hbox{\strut 75\textsuperscript{th} percentile)}}} \\\hline
		$F_b$ \hspace{2.3mm}  (\si{\milli\litre\per\min\per\gram})           & 1.41 & (0.94, 1.75)   \\
		$v_b$  \hspace{2.8mm}  (dimensionless)        & 0.06 & (0.05, 0.07)   \\
		$v_e$  \hspace{2.8mm}   (dimensionless)        & 0.12 & (0.1, 0.15)   \\
		$PS$  \hspace{1.6mm}    (\si{\milli\litre\per\min\per\gram})        & 0.49 & (0.33, 0.68)   \\
	\end{tabular}
	\label{mean and std}
\end{table}

\begin{figure*}[h!]
	\centering
	\includegraphics[width = .96\textwidth]{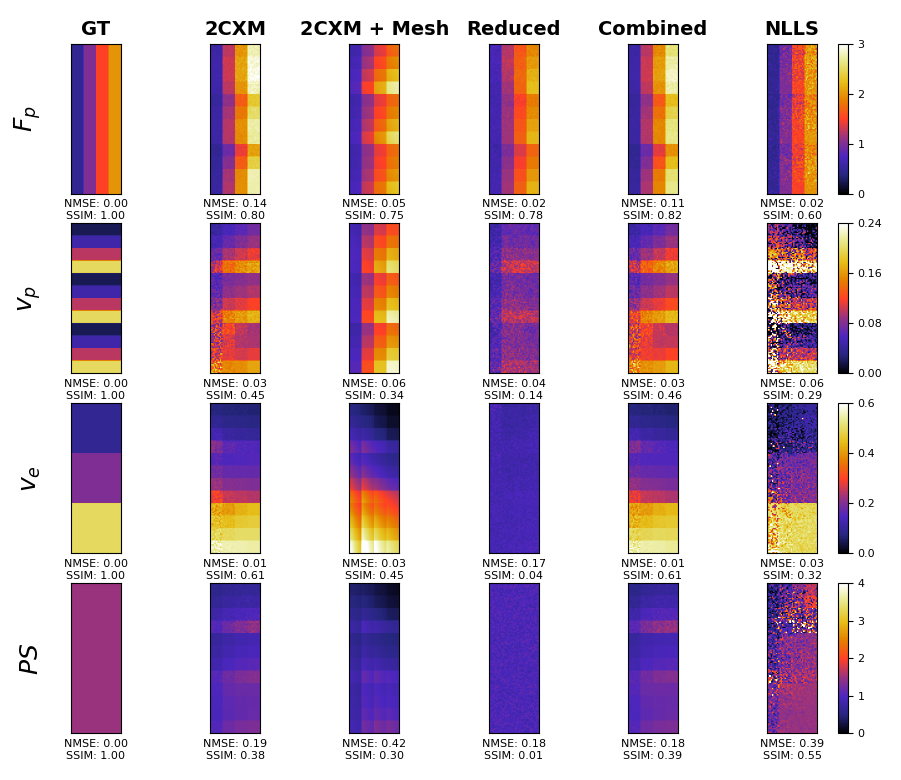}
	\caption{Results for the inference of the tracer-kinetic parameters taken from a single 2D slice of the DRO, performed by the five methods described in Section \ref{Dro_study}. Inference maps are provided for all four parameters, with the first columns denoting the ground truth maps. The next four columns present the results for the different PINN methodology, with the last row showing the results for the analytical solution. NMSE and SSIM values are provided for each map relative to the ground truth.}
	\label{DRO_PS1,5}
	\vspace{-11pt}
\end{figure*}

\begin{figure*}[h!]
	\centering
	\begin{subfigure}{.7\textwidth}
		\includegraphics[width=\textwidth]{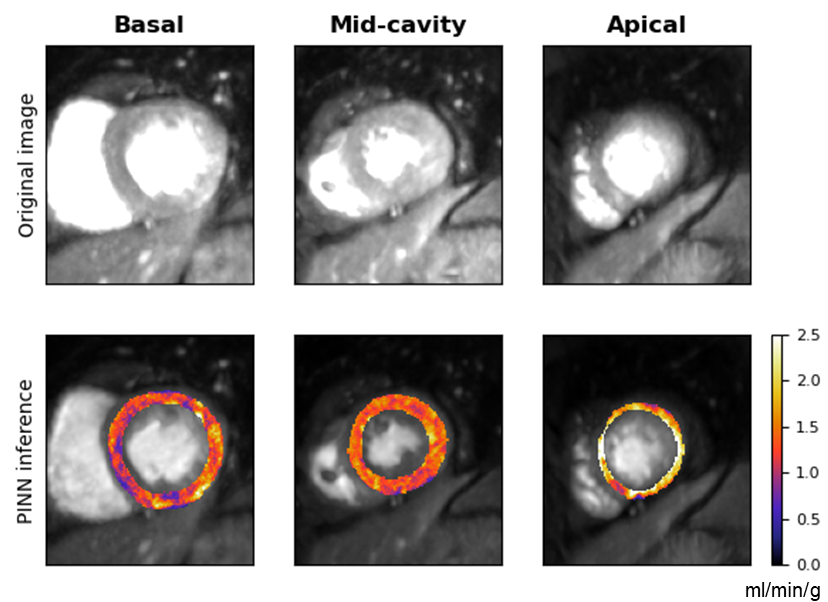}
		\hfill
	\end{subfigure}
	\hspace{10mm}
	\vspace{-3pt}
	~
	\begin{subfigure}{.7\textwidth}
		\includegraphics[width=\textwidth]{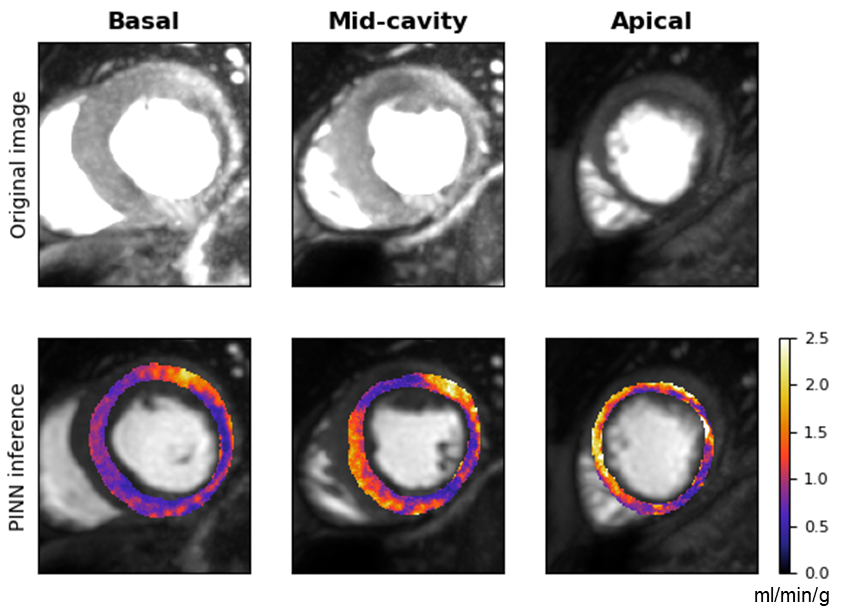}
		\hfill
	\end{subfigure}
	\vspace{-3pt}
	\caption{The estimated MBF maps for a patient (number 8) with no significant CAD (top) and patient (number 5) with CAD (bottom). The maps are shown under the corresponding MR images. The contrast of the MR images has been stretched to try to more clearly visualise the ischaemic regions.}
	\vspace{-10pt}
	\label{pt_results}
\end{figure*}

\section{Discussion}

The aim of this study was to demonstrate the feasibility of inferring tracer-kinetic parameters from DCE-MRI data using physics-informed neural networks. PINNs are used to solve supervised learning tasks while respecting the underlying physics of the problem, usually represented in the form of differential equations. This allows the training of universal function approximators to solve differential equations that are highly data-efficient due to the enforcement of the physical laws. This paradigm can be extended beyond solving differential equations to inference problems which considers the kinetic parameters to be inferred as trainable parameters of the PINN.

Since first introduced by Raissi et al.\cite{Raissi2019}, PINNs have been employed for both solving differential equations and estimating parameters across a wide range of physical problems \cite{Kadeethum2020,Tartakovsky2020}, including cardiac electrophysiological modelling \cite{Costabal2020} and blood flow modelling \cite{Kissas2020}. In this study, PINNs of different forms were developed and their performance was tested and compared to the traditional non-linear least squares fitting method in a simulated environment. The PINN approach was shown to outperform the NLLS solutions. The methods were further applied to patient data and yielded quantitative values close to the expected ranges. The MBF values are possibly lower than typically reported due to the high prevalence of ischaemia in these patients. The MBF values were also shown to discriminate well between normal and ischaemic tissue. To the best of our knowledge, this study presents the first use of PINNs for the inference of kinetic parameters in myocardial perfusion MR and DCE-MRI in general. Despite the fact that this is an initial proof-of-concept study, the promising results mean that it warrants further study which could lead to the adoption of PINNs as the method of choice for kinetic parameter estimation. Particularly since the adoption of quantitative MRI in clinical practice has been slow due to concerns about the accuracy and standardisation of methods. 

Based on the DRO study, the Combined PINN has the best overall performance with low NMSE and high SSIM as compared to the ground-truth kinetic parameters. This also shows the benefits of the flexibility of the PINN approach. It is seen that the Reduced PINN improves the results of the $F_p$ estimation as compared to the 2CXM PINN while the other three parameters suffer. However, a combination of the two models, in the Combined PINN, gives better results for all parameters. $PS$ is consistently the most difficult parameter to estimated and has high NMSE and low SSIM metrics. However, since the same trend is found with all methods it may be to do with the data sampling rather than the fitting. Stress myocardial perfusion MR acquisitions may not be long enough to measure the complete wash-out of the contrast agent and thus the full information regarding $PS$ may not be present in the data. Typical stress myocardial perfusion MR studies only report MBF values, however, as shown here, it is feasible to also assess the other kinetic parameters. It is possible that these parameters will give more in-depth insight to the microvasculature, a topic which is gaining increased attention in the literature \cite{Rahman2019, Rahman2020}.

While PINNs are based on tracer-kinetic models, they offer all of the flexibility deep learning and leverage the recent advances in the field. This includes the flexible optimisation algorithms and loss functions that may be combined and changed at will, and may be designed with intuition based on the governing physical law system. For example, in this study it was noted that the residual loss terms may be calculated for any time point $t$, as the physical equations should hold regardless of the time input. So, while the observed measurements are fit at a limited number of points with a set temporal resolution (limited by the image acquisition process), the residuals can be enforced at any time point. Another property of conventional tracer-kinetic model fitting is that parameters are estimated pixelwise. That is, the estimates in each pixel are computed independently and spatial dependencies are rarely considered. However, these spatial correlations are exploited more easily in the PINN framework as a single neural network predicts parameters for all pixels simultaneously. Additionally, the spatial coordinates can be included as an input to the PINN to regularise the solution. The benefits of the joint optimisation of all kinetic parameters is clearly evident in Fig. \ref{DRO_PS1,5} as the PINN approaches show much fewer outliers than the pixel-by-pixel fitting of the NLLS method. The use of the mesh coordinates as a further input to the PINN does not improve the overall performance but it does also appear to have some benefits. That is that the kinetic parameter estimates computed with the 2CXM + Mesh PINN are seen to preserve the structure in the DRO better than the other models and thus, this approach could warrant further investigation.

While, in this work, different loss functions were tested and different inputs to the PINN were tried, there is also a huge potential to exploit the flexible framework further. For example, it would be possible to change the input from the image-derived concentration curves to the acquired k-t space and to learn both the image reconstruction and kinetic parameter estimation, similar to Dikaios \cite{Dikaios2020}.

It has been demonstrated previously that it is possible to train deep neural networks to directly estimate kinetic parameters \cite{Scannell2019deep}. However, this used a training set with corresponding labelled kinetic parameters. The difficulty of this approach is that there is not a ground-truth available for the training labels and these were derived from a separate inference scheme. The benefit of PINNs over such purely data-driven approaches is that it does not require labelled training data.

An important limitation of the proposed approach to consider for the inference of kinetic parameters is the computational cost as each individual slice or patient requires the training of a neural network. The proposed PINN method requires approximately 1 hour per imaging slice utilizing an NVIDIA GeForce GTX 980M series GPU, as opposed to 3 minutes for the NLLS fitting. A possible solution to this could be the use of transfer learning. Using the assumption that most patients have similar kinetic parameters and concentration curves, a baseline PINN could be trained sequentially from a number of different patients and future patients could be processed by fine-tuning these weights rather than training from scratch. Initialising the weights from previously trained examples may also reduce spurious updates and increase robustness.

\section{Conclusion}
In this work, we have shown that it is feasible to perform myocardial perfusion quantification with physics-informed neural networks. Though the inference of kinetic parameters using PINNs is still slow and future work is required, this framework provides a high degree of flexibility and the initial results are promising. The methods for training neural networks are also still evolving at a high pace, making PINNs a good basis for future research. This may allow physics-informed machine learning to become an established method for tracer-kinetic modelling and quantitative MRI in general.

\section*{Declaration of Competing Interest} 
\noindent Marcel Breeuwer is an employee of Philips Healthcare. All other authors declare that they have no known competing financial interests or personal relationships.

\section*{Acknowledgments}

\noindent This work was supported by the King’s College London \& Imperial College London EPSRC Centre for Doctoral Training in Medical Imaging [EP/L015226/1]; Philips Healthcare; The Department of Health via the National Institute for Health Research (NIHR) comprehensive Biomedical Research Centre award to Guy’s \& St Thomas’ NHS Foundation Trust in partnership with King’s College London and King’s College Hospital NHS Foundation Trust; The NIHR Cardiovascular MedTech Co-operative; The British Heart Foundation [TG/18/2/33768]; and the Wellcome/EPSRC Centre for Medical Engineering [WT 203148/Z/16/Z].

\section*{Appendix A}
For a single pixel $j$, the constituent terms of the loss function $\mathcal{L}$ given in Equation \ref{Combined_loss} are:
\begin{align*}
		\mathcal{L}_{C}^j & = \frac{1}{N_{C}}\sum\limits_{i=1}^{N_{C}}(C_{myo}^j(t_i) - \hat{C}_{myo}^j(t_i; v_p^j, v_e^j, \theta))^2 \\
		& + \frac{1}{N_{C}}\sum\limits_{i=1}^{N_{C}}(C_{AIF}(t_i) - \hat{C}_{AIF}(t_i; \theta))^2 \tag{A.1} \\[10pt]
		\mathcal{L}_{r}^j & = \frac{1}{N_r}\sum\limits_{i=1}^{N_r}(r_p(t_i; \eta^j, \theta))^2 + \frac{1}{N_r}\sum\limits_{i=1}^{N_r}(r_e(t_i; \eta^j, \theta))^2 \\
		& + \frac{1}{N_r}\sum\limits_{i=1}^{N_r}(r_{myo}(t_i; \eta^j, \theta))^2  \tag{A.2} \\[10pt] 
		\mathcal{L}_{b}^j & =  \hat{C}_{p}^j(0; \theta)^2 + \hat{C}_{e}^j(0; \theta)^2 + \hat{C}_{myo}^j(0; v_p^j, v_e^j, \theta)^2 \\ 
		& + \hat{C}_{AIF}(0; \theta)^2  \tag{A.3} \\[10pt]
		\mathcal{L}_{reg}^j & = \frac{1}{N_r}\sum\limits_{i=1}^{N_r}(\min{(\hat{C}_{p}^j(t_i; \theta), 0)})^2 \\
		& + \frac{1}{N_r}\sum\limits_{i=1}^{N_r}(\min{(\hat{C}_{e}^j(t_i; \theta), 0)})^2. \tag{A.4}
\end{align*}
In these equations, $N_C$ denotes the total number of contrast agent concentration measurements derived from the DCE-MRI acquisition. $\hat{C}_{myo}^j(t_i; v_p^j, v_e^j, \theta)$ is the PINN predicted myocardial concentration at time $t_i$, $\hat{C}_{p}^j(t_i; \theta)$ is the PINN predicted plasma concentration, $\hat{C}_{e}^j(t_i; \theta)$ is the PINN predicted interstitial concentration, and $C_{AIF}(t_i; \theta)$ is the PINN predicted AIF, which both depend on the learned parameters of the PINN: $\theta$. $N_r$ is the number of collocation points, randomly chosen time-domain points for which the residuals of the ODEs are calculated. $\eta$ is the 2CXM parameters. 

\section*{Appendix B}
While the scale normalisation of $C_{AIF}$ and $C_{myo}$ cancels out from both sides of the ODEs, the standardisation of the time input from $t$ to $\hat{t}$ leads to Eq. \ref{ODEp}, \ref{ODEe}, and \ref{ODEmyo} being re-written as:

\begin{equation}
\begin{split}
\frac{1}{\sigma_t}v_p\frac{d\hat{C}_p}{d\hat{t}} & = F_p(\hat{C}_{AIF} - \hat{C}_p) + PS(\hat{C}_e - \hat{C}_p), \\
\frac{1}{\sigma_t}v_e\frac{d\hat{C}_e}{d\hat{t}} & = PS(\hat{C}_p - \hat{C}_e), \\
\frac{1}{\sigma_t}\frac{d\hat{C}_{myo}}{d\hat{t}} & = F_p(\hat{C}_{AIF} - \hat{C}_p).
\end{split}
\label{adjusted_ODEs} \tag{B.1}
\end{equation}
\setcounter{figure}{0}
\renewcommand{\thefigure}{C\arabic{figure}}
\begin{figure*}
\section*{Appendix C}
	\centering
	\includegraphics[width = 0.66\linewidth]{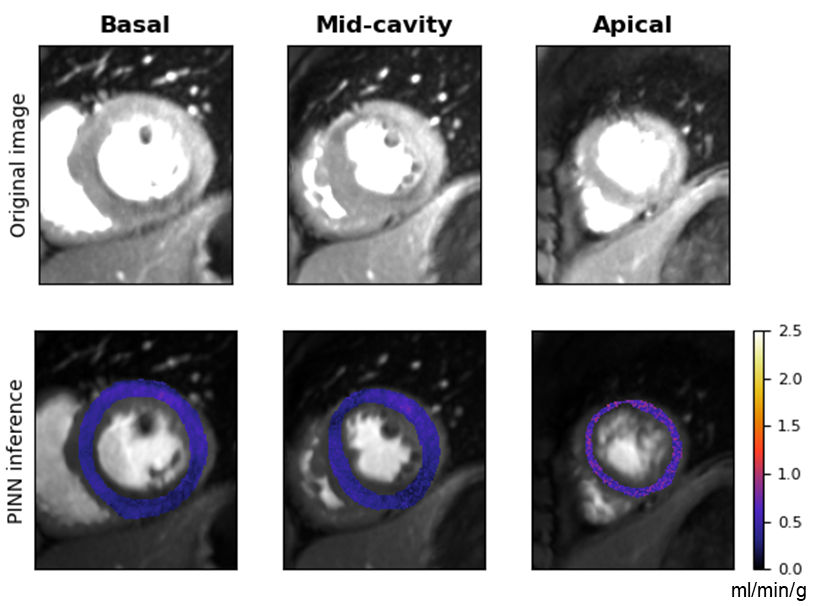}
	\caption{The estimated MBF maps and corresponding MR images for patient number 1.}
	\label{pt1}
\end{figure*}
\begin{figure*}
	\centering
	\includegraphics[width = 0.66\linewidth]{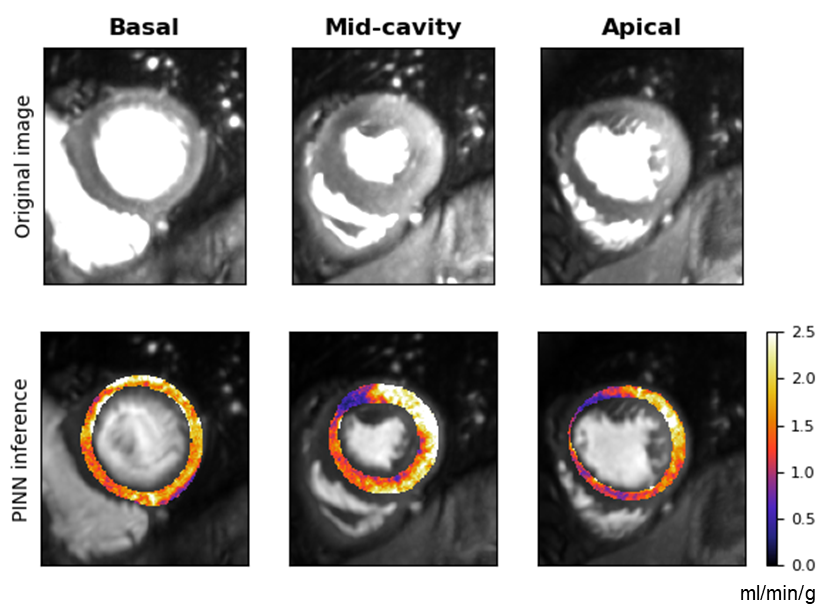}
	\caption{The estimated MBF maps and corresponding MR images for patient number 2.}
	\label{pt2}
\end{figure*}
\begin{figure*}
	\centering
	\includegraphics[width = 0.66\linewidth]{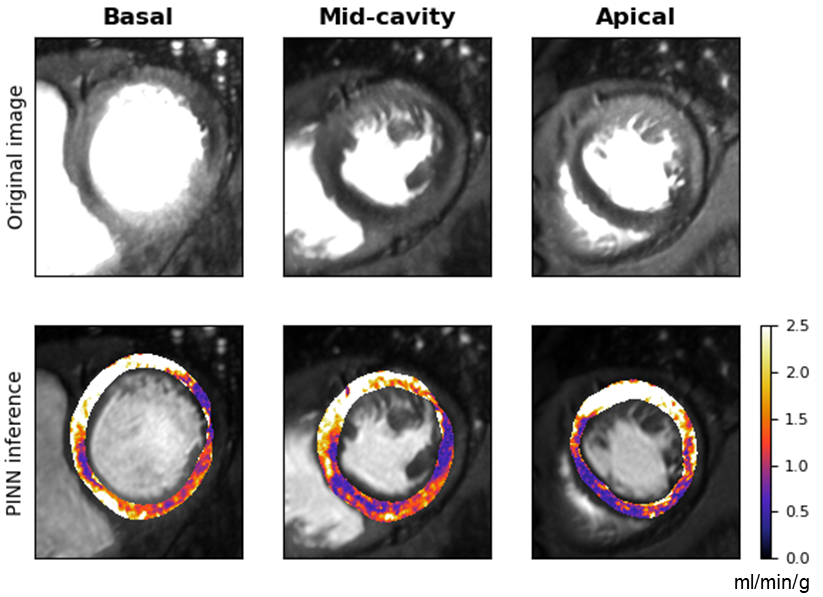}
	\caption{The estimated MBF maps and corresponding MR images for patient number 3.}
	\label{pt3}
\end{figure*}
\begin{figure*}
	\centering
	\includegraphics[width = 0.66\linewidth]{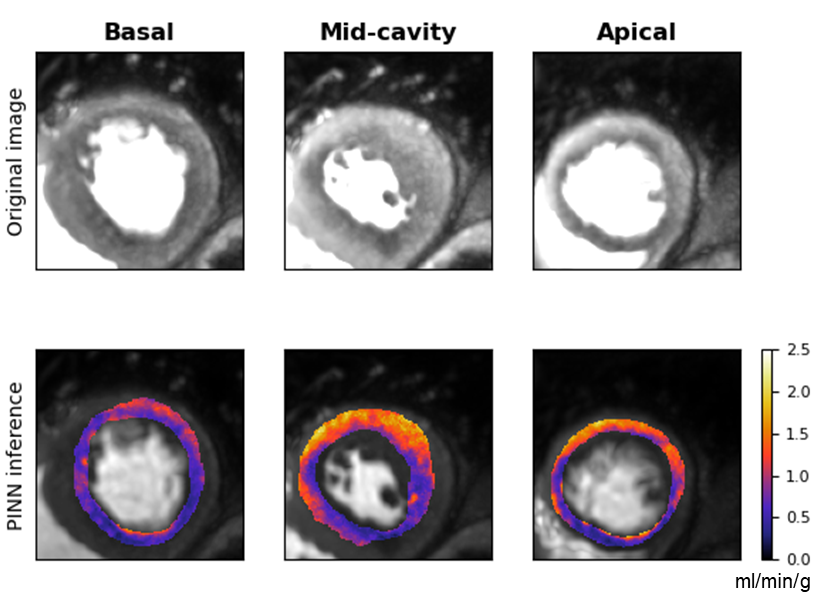}
	\caption{The estimated MBF maps and corresponding MR images for patient number 4.}
	\label{pt4}
\end{figure*}
\begin{figure*}
	\centering
	\includegraphics[width = 0.66\linewidth]{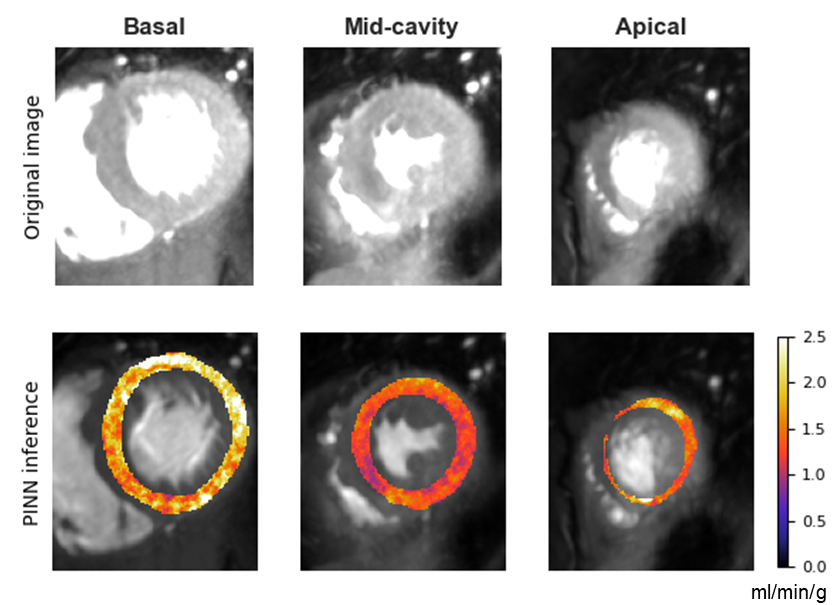}
	\caption{The estimated MBF maps and corresponding MR images for patient number 6.}
	\label{pt6}
\end{figure*}
\begin{figure*}
	\centering
	\includegraphics[width = 0.66\linewidth]{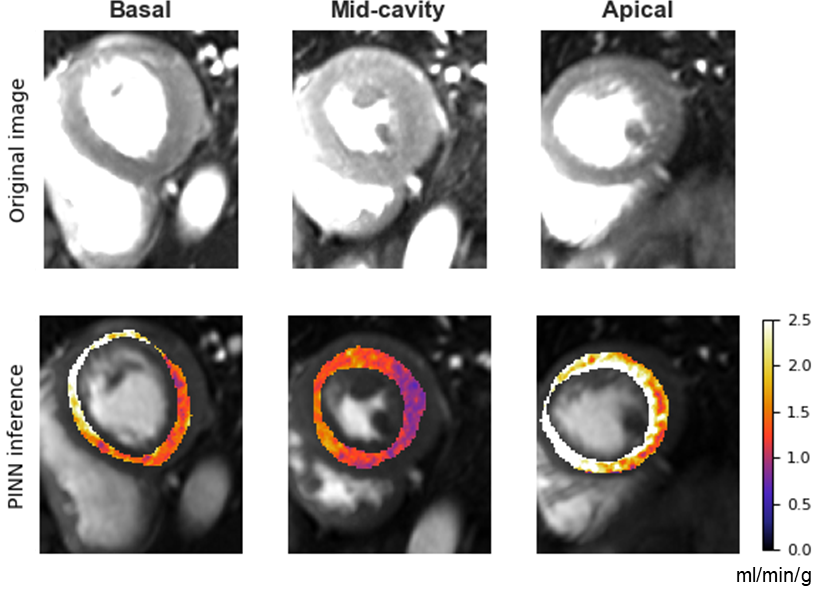}
	\caption{The estimated MBF maps and corresponding MR images for patient number 7.}
	\label{pt7}
\end{figure*}

\onecolumn{
	\bibliographystyle{abbrv}
	\bibliography{new_refs}
}

\end{document}